\documentclass{article}

\RequirePackage[OT1]{fontenc}
\RequirePackage{amsthm,amsmath,geometry,amsfonts}
\RequirePackage{natbib}
\RequirePackage[colorlinks,citecolor=blue,urlcolor=blue]{hyperref}

\usepackage{longtable}

\usepackage{array}
\newcolumntype{L}[1]{>{\raggedright\let\newline\\\arraybackslash\hspace{0pt}}m{#1}}
\newcolumntype{C}[1]{>{\centering\let\newline\\\arraybackslash\hspace{0pt}}m{#1}}
\newcolumntype{R}[1]{>{\raggedleft\let\newline\\\arraybackslash\hspace{0pt}}m{#1}}

\usepackage{float, graphicx, enumitem, caption}

\usepackage[nokeyprefix]{refstyle}
\usepackage{varioref}
\usepackage{xr-hyper}
\usepackage{hyperref}

\numberwithin{equation}{section}

\title{Adjusting for Unmeasured Spatial Confounding with Distance Adjusted Propensity Score Matching
}
\author{Georgia Papadogeorgou, Christine Choirat, Corwin Zigler \\[10pt]
	\normalsize Harvard T.H. Chan School of Public Health}
\usepackage{fancyhdr}
\date{}

\begin{document}

\maketitle

\begin{abstract}
{Propensity score matching is a common tool for adjusting for observed confounding in observational studies, but is known to have limitations in the presence of unmeasured confounding.  In many settings, researchers are confronted with spatially-indexed data where the relative locations of the observational units may serve as a useful proxy for unmeasured confounding that varies according to a spatial pattern.  We develop a new method, termed Distance Adjusted Propensity Score Matching (DAPSm) that incorporates information on units' spatial proximity into a propensity score matching procedure. We show that DAPSm can adjust for both observed and some forms of unobserved confounding and evaluate its performance relative to several other reasonable alternatives for incorporating spatial information into propensity score adjustment. The method is motivated by and applied to a comparative effectiveness investigation of power plant emission reduction technologies designed to reduce population exposure to ambient ozone pollution. Ultimately, DAPSm provides a framework for augmenting a ``standard'' propensity score analysis with information on spatial proximity and provides a transparent and principled way to assess the relative trade offs of prioritizing observed confounding adjustment versus spatial proximity adjustment.}
\end{abstract}

\textit{keywords:} {propensity score matching, spatial confounding, unobserved confounding}

\section{Introduction}
Methods based on propensity score matching are widely used to estimate causal effects with observational data. Such methods rely crucially on the assumption of no unmeasured confounding. In settings of spatially-indexed data, unobserved confounders may exhibit a spatial pattern, inviting the use of spatial information to serve as proxy for similarity of units with respect to unmeasured confounding factors.  Methods for confounding adjustment with spatially-indexed data have been most often considered in the context of regression adjustment, as in \cite{Paciorek2010} and in related work that does not target confounding adjustment per se, but has been used for modeling spatially correlated residuals via spatial random effects \citep{Hodges2010, Lee2010, Lee2015a, Chang2013, Congdon2013}.

In this paper, we unite the use of spatially-indexed data with propensity score matching while preserving the most salient benefits of using propensity scores. These benefits include the explicit comparison of treatments or policy interventions to estimate policy-relevant estimands such as the Average Treatment Effect on the Treated, as well as the oft-cited virtues of propensity score analysis related to the hypothetical ``design'' of a randomized study, for example, the ability to check observed covariate balance and overlap \citep{Rubin2008}. Augmenting such benefits with the notion that geographically closer units may exhibit similar unmeasured confounding profiles presents a methodological challenge.

\begin{figure}[t]
	\vspace{20pt}
	\begin{center}
		\includegraphics[width = \textwidth]{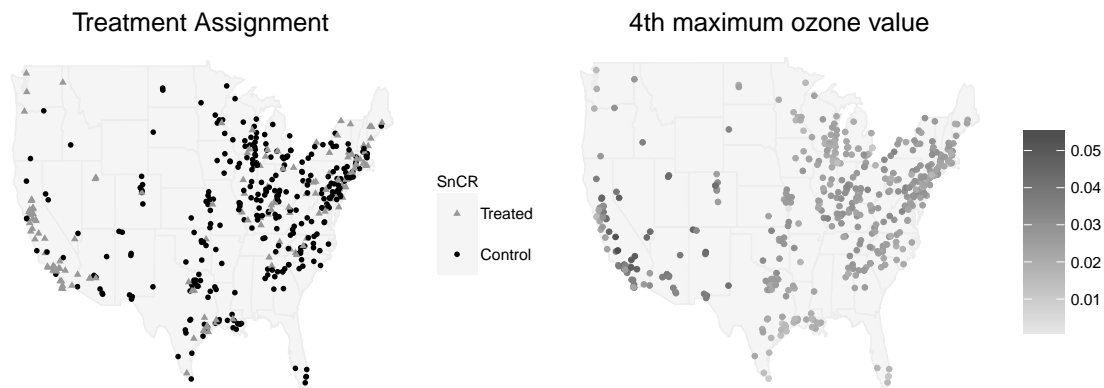}
		\caption{Map of facilities, colored by whether they are treated (yellow) or control (red), and map of ozone concentration surrounding power plants.}
		\label{fig:data_map}
	\end{center}
\end{figure}

The methods here are motivated by the threat of unmeasured spatial confounding that arises in studies of air pollution, where complex climatological and atmospheric processes are known to vary spatially and have strong associations with ambient air pollution, but are often unmeasured.  For example, consider ambient ozone pollution, which has been previously linked to adverse health outcomes \citep{Bell2004, Jerrett2009}. A variety of regulatory strategies in the U.S. are designed to reduce ambient ozone pollution through incentivizing power-generating facilities (i.e., ``power plants'') to reduce emissions of nitric oxide and nitrogen dioxides (NO$_x$). When combined with sunlight and in the presence of available volatile organic compounds, NO$_x$ emissions initiate atmospheric chemical reactions to form ambient ozone pollution \citep{Allen2003}. What's more, regions where conditions tend to encourage the formation of ozone might be more likely to impose stricter rules on NO$_x$ emissions. Thus, evaluating the effectiveness of emission-control strategies installed at power plants is met with the challenge that complete data on all relevant climatological, atmospheric, and regulatory confounders is almost never available but are expected to vary spatially. The goal of this paper is to employ a matching procedure anchored to the propensity score to investigate whether, among coal or natural gas power plants, installation of selective catalytic or selective non-catalytic (SCR/SNCR) NO$_x$ emission control technologies is more effective than alternatives for reducing ambient ozone. The treatment assignment and outcome of interest are depicted in Figure \ref{fig:data_map}. Propensity scores are particularly useful for this type of policy evaluation because of the ability to adjust for confounding without strong reliance on a parametric model and the ability to empirically assess covariate balance and overlap.  However, unmeasured spatial confounding presents a strong threat to the validity of a standard propensity-score analysis.

To confront these challenges, we present a new methodology, termed Distance Adjusted Propensity Score Matching (DAPSm) which incorporates information from spatially-indexed data with the known virtues of propensity score matching. DAPSm incorporates observations' spatial proximity into a matching procedure designed to adjust for observed confounders while adjusting for unmeasured spatial confounders by emphasizing, to varying degrees governed by a tuning parameter, the spatial proximity of matches. The central challenge of incorporating spatial proximity into propensity score matching is that proximity is a relative measure between  two units, not a unit-specific measure like a confounder or the propensity score itself.

DAPSm shares important commonalities with the recently-proposed work of \cite{Keele2015} in that both are matching methods that aim to leverage spatial proximity of units.  We evaluate DAPSm relative to the method of \cite{Keele2015} throughout, but note here that, despite similar conceptual goals, these methods are not directly comparable. The most salient difference has to do with reliance on the propensity score; DAPSm combines spatial proximity with propensity scores, whereas \cite{Keele2015} provides an integer programing method that matches directly on covariates (i.e., not using propensity scores).  Our goal here is not to investigate the relative merits of exact vs. propensity score matching, but rather to isolate features related specifically to methods' account of spatial confounding.  DAPSm offers a tuning parameter governing the relative prioritization of observed covariate distances (measured through similarity of propensity score estimates) and spatial proximity.  \cite{Keele2015} entails a tuning parameter that governs the trade off between spatial proximity of matches and the number of matches selected within a certain tolerance of observed covariate balance.

Both DAPSm and the method of \cite{Keele2015} are evaluated in a simulation study alongside several other reasonable alternatives for incorporating spatial information into propensity score analysis.  The methods are then deployed to compare the effectiveness of SCR/SNCR, relative to other strategies, for reducing NO$_x$ emissions and ambient ozone measured across 473 power plants and 921 air pollution monitoring locations in the United States. Ultimately, we show that incorporating spatial information in the matching can lead to substantively different conclusions when evaluating interventions on spatially-indexed observational units.


\section{Notation, estimand of interest, and outline of propensity score matching}

Let $Z_i$ denote the indicator of whether the $i^{th}$ of $n$ observations is subject to treatment, for example, the indicator of whether a power plant is treated with SCR/SNCR ($Z_i = 1$) or not $(Z_i = 0)$.  Let $Y_i$ be a continuously-scaled outcome, for example, ambient ozone concentration in the area surrounding power plant $i$.

Each unit or observation is assumed to have two potential outcomes \citep{Rubin1974}, one under each value of the treatment. We denote $Y_i(0), Y_i(1)$ as the potential outcome of ambient ozone concentration in the area around unit $i$ under value of $z = 0, 1$ respectively. Assuming that the indexing of the observations is done at random, the index $i$ is suppressed. Interest often lies in the estimation of the average treatment effect in the treated population, defined as
$ ATT = E[Y(1) - Y(0) | Z = 1] .$

Among the assumptions required to estimate the ATT with observational data is that of ``ignorable treatment assignment'', stating that observed covariates are sufficient to adjust for confounding of the treatment-outcome relationship. More formally, let $\textbf{C}$ be a minimal set of confounding variables such as power plant characteristics, weather and atmospheric variables, and area-level demographics. The assumption of ignorability can be stated as:
\begin{equation}
Y(z) \amalg Z | \textbf{C},
\label{eq:counter_true}
\end{equation}
under which the ATT can be estimated with observed-data comparisons between outcomes on treated and untreated units, conditional on $\textbf{C}$. Since $\textbf{C}$ is assumed minimal, the ignorability assumption in (\ref{eq:counter_true}) does not hold for any strict subset of $\textbf{C}$, implying that observed-data comparisons will not estimate the ATT when conditioning on a strict subset of \textbf{C}.

As the dimensionality of $\textbf{C}$ increases, investigators often use the propensity score to condense the information in \textbf{C} into a ``balancing score'' that can be used to adjust for confounding when comparing treated and untreated units. The propensity score is defined as the conditional probability of receiving treatment given the covariates, $P(Z_i = 1 | \textbf{C}_i)$.
The balancing property of the propensity score \citep{Rosenbaum1983} implies that the ignorability assumption in (\ref{eq:counter_true}) can be translated to
$
Y(z) \amalg Z | P(Z = 1 | \textbf{C}).
$

An overview of the various ways in which propensity scores can be used for confounding adjustment can be found in \cite{Stuart2010}, but we discuss methods in the context of 1:1 matching without replacement using a caliper.  Such a procedure uses propensity score estimates to match one treated unit to one control unit with a similar propensity score estimate. A threshold, called a ``caliper,'' can be used to avoid matching observations with insufficiently similar propensity scores.  For example, specifying a caliper of 0.1 prevents the matching of any two observations with propensity scores that differ by more than 0.1 standard deviations of the propensity score distribution. Matching produces a data set of matched treated and control observations with similar propensity score distributions, and thus more similar distributions of the covariates in $\textbf{C}$ and a treatment indicator that is, under ignorability, unconfounded.
The resulting matched data set can be used to estimate the ATT provided that all elements of $\textbf{C}$ are observed and used to construct the propensity score.

However, it is often the case with observational data that the vector of true confounders $\textbf{C}$ can be partitioned into two categories, $\textbf{C} = (\textbf{X}, \textbf{U})$, where \textbf{X} denotes the confounders available in the observed data, and \textbf{U} denotes confounders that are unobserved.
In the presence of unobserved confounders $\textbf{U}$, the ignorability assumption in ({\ref{eq:counter_true}}) cannot be satisfied by conditioning solely on the observed \textbf{X}, and the treatment effect is not identifiable from the data.

In many settings it is expected that some elements of \textbf{U} vary spatially so that locations that are geographically close are similar with regard to \textbf{U}.  In this sense, the notion of prioritizing spatial proximity of matches has points of contact with the notion of a spatial ``bandwidth'' in a geographic regression discontinuity design (such as that in \cite{Keele2015}), where only observations within the bandwidth are regarded as comparable on observed (and unobserved) factors.  The method outlined below regards $\mathbf{U}$ as unmeasured variables with a distribution over the whole geography of interest that is continuous as a function of space, with closer observations having more similar \textbf{U}.

\section{Distance Adjusted Propensity Score Matching}
We propose a procedure that is anchored to the propensity score for matching on observed confounders, but augments confounding adjustment by incorporating spatial (geographical) information as a proxy for unobserved spatial variables, $\textbf{U}$, such as weather and atmospheric conditions.
In the presence of such $\textbf{U}$, prioritizing matched units that are geographically close to each other could yield better covariate balance on all $\textbf{C} = (\textbf{X}, \textbf{U})$, thus (approximately) recovering the ignorability assumption and reducing bias of causal estimates. Formally, the variables $U \in \textbf{U}$ are such that $\forall \epsilon > 0$ and point $s_0$ in the geography of interest $\exists \mathcal{N}_\epsilon(s_0)$ open set including $s_0$ such that $|U(s) - U(s_0)| < \epsilon$, $\forall s \in \mathcal{N}_\epsilon(s_0)$.

We define the Distance Adjusted Propensity Score (DAPS) as a new quantity for identifying good matches between treated and control units. In contrast to the propensity score which has a value for each unit, the DAPS is defined for every $(i,j)$ pair of treated, control observations. Specifically, for treated unit $i$ and control unit $j$, the DAPS combines propensity score estimates and relative distances to define:
$ DAPS_{ij} = w * | PS_i - PS_j |  + (1-w) * Dist_{ij}, $
where $w \in [0, 1]$, $PS_i, PS_j$ are propensity score estimates from modeling the treatment conditional on the observed confounders, and $Dist_{ij}$ is a distance measure capturing the proximity of units i, j. DAPS is a weighted average of the propensity score difference used in ``standard'' propensity score matching and a measure of the distance between treated-control pairs.  Therefore, it is a transparent measure of similarity between treated and control units, with an $(i,j)$ pair having small DAPS$_{ij}$ regarded as comparable on the basis of a combination of propensity score difference and spatial proximity. 


\subsection{Choosing the weight $w$}
Setting $w = 1$ corresponds to setting DAPS equal to the absolute propensity score difference, and similarity of treated and control units is based solely on the observed confounders, without regard to spatial proximity. Setting $w = 0$ ignores $\mathbf{X}$, defining similarity of units based solely on distance.  In practice, $w$ could be specified in the range $[0,1]$ depending on contextual prioritizaiton of observed confounding and the threats due to any suspected unobserved spatial confounding, with values closer to 0 for settings where unobserved spatial confounding is of particular concern. Data-driven procedures, such as the one described in Section \ref{sec:choose_w} can be useful in choosing a $w$.

\subsection{Choosing the distance measure}
The quantity $Dist_{ij}$ could be specified in many ways to quantify spatial proximity of units $i$ and $j$.  A natural distance measure is the geographical distance between units $i, j$. A key consideration in choosing a distance measure is that its scale must be made comparable to that of the propensity score to ensure that one quantity does not arbitrarily dominate the calculation of DAPS. Since the absolute propensity score difference of two units can vary across the range $[0,1]$, the distance measure should also vary between 0 and 1, or on a range similar to the range of estimated propensity score differences. (Alternatively, instead of standardizing $Dist_{ij}$, one could scale $w$.)

One distance measure we consider is the standardized Euclidean distance (for simulations) or the standardized geo-distance (for the application). Specifically, if $i \in S_t = \{ 1, 2, \dots, N_t\}$ is a treated unit, and $j \in S_c = \{1, 2, \dots, N_c\}$ is a control unit, the standardized distance of $i, j$ is defined as:
\begin{equation}
Dist_{ij} = \frac{d_{ij} - {\min}_{TC} d }{ \max_{TC} d - \min_{TC} d},
\label{eq:SD1}
\end{equation}
where $d_{ij}$ is the Euclidean (or geo) distance between $i, j$, and $\min_{TC} d, \max_{TC} d$ are the minimum and maximum distances of all the treated-control pairs.

Other choices of distance measure can also arise in practice. For example, only permitting matches within certain boundaries (e.g., within states) corresponds to setting $Dist_{ij}=\infty$ for $i,j$ located in different states. Appendix \ref{app_sec:sd2} presents an alternative definition relying on the empirical CDF of treated-control pairwise distances.

\subsection{Selecting matches}
\label{sec:algorithm}
We provide an R package that performs matching based on DAPS using an optimal or a greedy algorithm. The optimal algorithm uses the \texttt{optmatch} R package, and the greedy algorithm is described in Appendix \ref{app_sec:greedy}.
\cite{Gu1993} found that optimal matching performed better than greedy matching in returning matched pairs with small Mahalanobis covariate distance, but returned similarly balanced matched data sets.

\subsection{Specifying Calipers}

In DAPSm, a caliper can be defined as the number of DAPS standard deviations beyond which a value of DAPS is deemed too large to produce an appropriate match. In this situation, a treated-control pair cannot be matched if the corresponding DAPS of the pair is larger than the caliper. That is, the caliper is directly applied to the entire DAPS quantity.

Calipers could be alternatively defined to pertain separately to each component of DAPS.  For example, one type of caliper could prevent any match with propensity score difference exceeding some threshold regardless of DAPS value, with an analogous caliper defined only for distance.

Note that when a caliper is not used, there is an equivalence between DAPSm with $w = 1$ and standard 1-1 nearest neighbor propensity score matching. When a caliper is specified these procedures may not be exactly equivalent due to the definition of the caliper for the two procedures. Standard matching uses the standard deviation of propensity score estimates, while DAPSm uses the standard deviation of DAPS  $ \overset{w = 1}{ = }| \text{PS difference} | $.

\subsection{Data-driven choice of $w$}
\label{sec:choose_w}

In DAPS, there is a transparent interplay between distance of observed covariates (as measured through the difference in the propensity score estimates) and distance of matched pairs. Automated data-driven procedures may be useful for selecting an appropriate value of $w$.
We implement an automated procedure that re-calculates DAPS and performs matching across a range of possible $w$. As $w$ increases, balance of the observed covariates can be assessed, and the smallest value of $w$ that maintains the absolute standardized difference of means (ASDM) of the observed confounders below a pre-specified cutoff is used.  A different balance criterion can also be used. This choice of $w$ assigns the largest possible weight to proximity (and, by extension, to the unmeasured spatial confounders), while still maintaining balance of the observed confounders.

Even though this choice of $w$ is such that it ensures observed covariate balance with respect to a specific criterion, $w$ can be specified alternatively if subject-matter knowledge is available on an unmeasured spatial confounder. For example, there may be a known but unmeasured confounder (e.g., volatile organic compounds, baseline NO$_x$ emissions  in an air pollution study) that is regarded as more important than any measured variable. The value of $w$ could be chosen such that DAPSm prioritizes spatial proximity of matched pairs to maximize the chance of balancing the unmeasured spatial confounder, even at the cost of balance on observed covariates. Ability to make such a judgment transparently is a key feature of DAPSm.

\section{Simulation and Comparison with Alternatives}
\label{sec:sims}

We conduct a simulation study to explore the performance of DAPSm and several reasonable alternatives for incorporating spatial information, with a focus on how different methods perform across a variety of unmeasured spatial confounding settings, as dictated by the spatial surface of simulated unmeasured confounding. We evaluate methods with respect to mean squared error (MSE) of ATT estimates, balance of observed and unobserved confounders, and number of matches. Data are simulated across the locations of 800 power generating facilities to represent a realistic spatial patterning of units reflecting that of the study of power plant emissions and ozone.  Specifically, for each simulated data set, each of 800 fixed locations are simulated to have one unmeasured confounder $U$ generated as a Gaussian Process with Mat\'ern correlation function, four observed confounders $X_i, i = 1, 2, 3, 4$ uncorrelated with $U$, binary treatment $Z$, and continuous outcome $Y$. The specifics of the data generating mechanism can be found in Appendix \ref{app:sims_generate}.

The Mat\'ern correlation function of the spatial confounder is governed by two parameters, the smoothness, $\nu$, and range, $r$. The range, $r$, measures how quickly the correlation of $U$ between two locations decays with distance. When $\nu$ is small, the spatial process is rough, and when it is large the process is smooth. See \cite{Minasny2005} for a detailed description. Appendix \ref{fig:Usurf} shows four generated surfaces of a spatial variable with Mat\'ern correlation function for combinations of small and large values of smoothness and range.

The situation presented here assumes that $U$ is uncorrelatd with \textbf{X} to highlight the impact of completely unobserved spatial confounders. Situations where $U$ is simulated to have correlation with \textbf{X} produce similar results with less pronounced gains of incorporating spatial information.

\subsection{Methods for Comparison with DAPSm}

We consider alternative approaches belonging to two general strategies for incorporating spatial information with propensity scores: a) incorporating spatial information in the matching procedure, and b) incorporating spatial information in the propensity score estimates themselves.  The former methods estimate the propensity score using $\mathbf{X}$, then perform matching based in part on distance, as done in DAPSm. The latter methods estimate propensity scores that vary according to a spatial pattern by construction, then matches on these ``spatial propensity scores''. After matching is performed, ATT estimates are acquired through a difference in means of the matched pairs. Further regressions adjustments could be performed in practice.

The previously described method of \cite{Keele2015} is one method that incorporates spatial information in the matching. Even though \cite{Keele2015} advocate for matching directly on covariates, in the simulation study this method was implemented performing exact matching on 5 categories of the propensity score, such that any difference in performance could be attributed solely to the methods' ability to adjust for unmeasured spatial confounding. Simulation results for this method implemented to match directly on covariates are shown in Appendix \ref{app_sec:sims_Keele}.

We further considered another method that incorporates spatial information into the matching procedure, which we refer to as ``Matching within Distance Caliper". For this method, a distance caliper is chosen as the maximum distance of potential matched pairs. Within the distance caliper, matching is performed based solely on the propensity score estimated with $\mathbf{X}$. A caliper on the propensity score can be used in addition.

Methods that incorporate spatial information into the propensity score estimates  include  parametric and non-parametric incorporation of spatial information in the treatment assignment model. A simple approach is the introduction of fixed effects for locations' latitude and longitude coordinates in the propensity score, in addition to the observed covariates. We refer to this as ``Na\"ive with Coordinates''. A more flexible extension is to use Gradient Boosting Models (GBM; \cite{Friedman2001}) to estimate the propensity score, including the coordinates and the observed covariates $\textbf{X}$. Estimation of the model is performed using the \textit{gbm} R package \citep{Ridgeway2007}.

While the ``Na\"ive with Coordinates'' and GBM approaches are not spatial methods per se, an alternative approach is to augment the propensity score model with a spatial random effect, as implemented using the \textit{spBayes} R package \citep{Finley}. Specifically, the propensity score is estimated by fitting 
$
P(Z = 1 | \textbf{X}) = f(\textbf{X}, W; \theta_W), \ \  W \sim GP(0, C),
$
where $C = C(\lambda)$ is the spatial random effect correlation matrix with parameters $\lambda$. Such approach was not pursued in detail here due to its computational intensity and its poor performance in initial investigations.

We compare the propensity score matching with spatial information methods to the gold standard which uses the data generating outcome model, and the gold standard propensity score (``Gold PS'') which uses the true propensity score model conditional on $\mathbf{X}$ and $U$. Finally, the na\"ive approach performs propensity score matching using estimates from a model solely on the observed confounders $\mathbf{X}$.

All methods are implemented with 1-1 nearest neighbor optimal matching without replacement. For DAPSm, we present results for the definition of standardized distance defined in (\ref{eq:SD1}). For GBM, we considered 3$^{rd}$ degree interactions. Results for Matching within Distance Caliper are presented with the distance caliper equal to the 10$^{th}$ percentile of pairwise treated-control distances (the method indicated sensitivity to the choice of distance caliper, other specifications were considered, but are not shown here). The method of \cite{Keele2015} was implemented across a range of values for $\lambda$, representing different compromises between the number of returned matches and the distance between matched pairs.  As \cite{Keele2015} do not provide specific guidance on the selection of $\lambda$, we present results for two values meant to represent two different points in the space of compromises between distance and the number of matches: $\lambda = 0.382$, the median pairwise distance of treated and control units (as done in \cite{Keele2015}), and  $\lambda = 0.05$ which was determined in simulation to yield fewer matches and lower MSE for this range of simulation scenarios. For implementing DAPSm, $w$ was chosen based on the algorithmic procedure described in section \ref{sec:choose_w}.


\subsection{Simulation Results}

\begin{figure}[t]
	\centering
	\includegraphics[width = 0.93\textwidth]{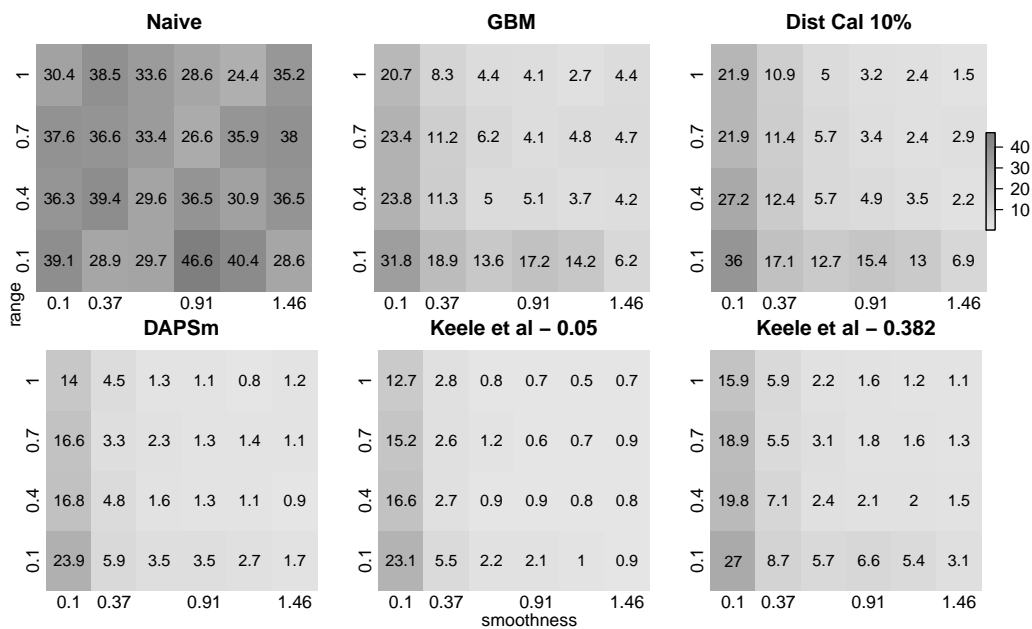}
	\caption{Estimates of relative mean squared error over 100 simulated data sets for each specification of smoothness $\nu$ (x-axis) and range $r$ (y-axis) for the Mat\'ern correlation function of the unobserved confounder. The baseline MSE corresponds to the Gold PS. Printed values are rounded to the first decimal.}
	\label{fig:mse_irr}
\end{figure}

Figure \ref{fig:mse_irr} shows the relative MSE of the effect estimates calculated with a subset of the methods with respect to the Gold PS and for different specifications of smoothness and range of $U$. MSE is calculated over the subset of simulated data sets for which each method returned matches. Table \ref{table:dropped_units} describes the percentage of simulated data sets for which no matching was achieved for each of the methods.
As expected, the na\"ive approach has the highest relative MSE ranging from 24.4 to 46.6. Relative MSE for the gold standard varied from 0.16 to 0.32, indicating that specifying the correct outcome model is more efficient than using the correctly specified propensity score. These approaches did not indicate patterning when varying the spatial structure. Relative MSE for the Na\"ive with coordinates ranged from 6 to 41.3, and indicated similar patterning as the other spatial methods with respect to the smoothness and range of $U$, but performed worse in terms of MSE than its more flexible form (GBM) and is emitted from Figure \ref{fig:mse_irr}.

For all methods incorporating spatial information, relative MSE decreases as the surface gets smoother (larger values of smoothness $\nu$) or the spatial correlation remains positive at longer distances (larger values of range $r$). Similar results are observed for the absolute bias.
Among the methods considered based on propensity scores, DAPSm had the lowest MSE across all specifications of range and smoothness, apart from the method of \cite{Keele2015} when $\lambda = 0.05$ was chosen such that it reduces simulation-based MSE compared to $\lambda = 0.382$.
Results for the method of \cite{Keele2015} implemented to match directly on the observed covariates, instead of the propensity score, can be found in Appendix \ref{app_sec:sims_Keele}, and showed lower relative MSE than the methods presented here.

\begin{figure}[p]
	\centering
	\includegraphics[width = \textwidth]{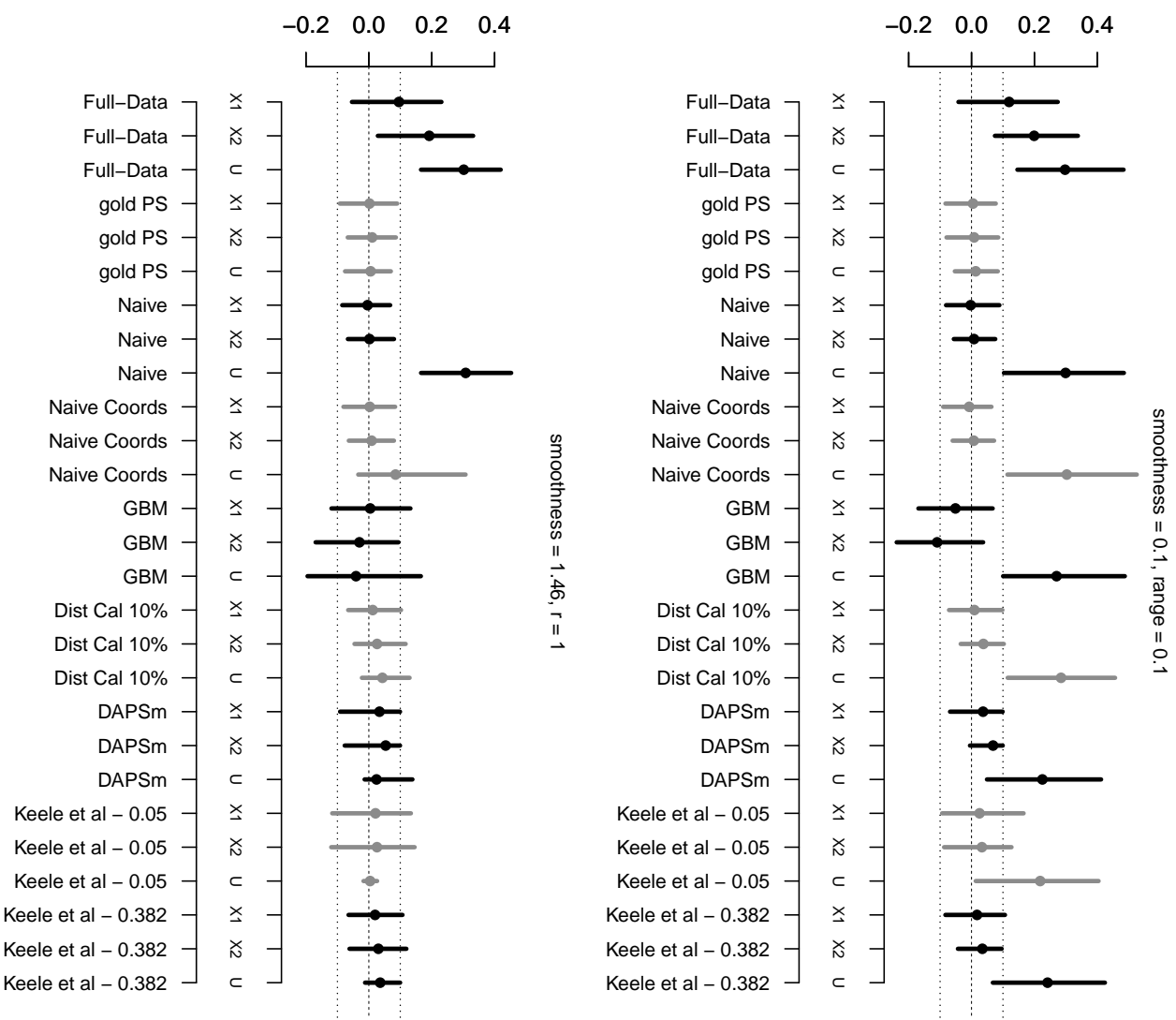}
	\caption{Average and (2.5\%, 97.5\%) intervals of standardized difference of means of $(X_1, X_2, U)$ for the scenarios where $\nu = r = 0.1$ (top), and $\nu = 1.46, r = 1$ (bottom) over 100 Monte Carlo simulations.}
	\label{fig:balance}
\end{figure}

We also evaluated methods with respect to the balance of observed and unobserved covariates. Figure \ref{fig:balance} shows the standardized difference of means of $X_1, X_2, U$ (balance of $X_3, X_4$ was similar to the balance of $X_1$, $X_2$) for the scenarios where $\nu = r = 0.1$ (rough uneven surface), and $\nu = 1.46, r = 1$ (smooth surface). First, the full data ASDM shows that all variables were imbalanced in most simulated data sets. Using the correctly specified propensity score model (Gold-PS) achieved balance of all confounders at the 0.1 cutoff. The na\"ive approach does not incorporate any spatial information, and the unobserved confounder remains imbalanced. Incorporating coordinates in the estimation of the propensity score improves on balancing $U$, especially in smoother surfaces. Matching within Distance Caliper performed similarly to Na\"ive with coordinates in the rough surface. In smooth surfaces Matching within Distance Caliper performed well in balancing both \textbf{X} and $U$, although the balance of $U$ is sensitive to the choice of the distance cutoff. The GBM approach exhibited poor balance for all covariates in both scenarios. DAPSm with weight $w$ chosen as described in section \ref{sec:choose_w} balanced all observed covariates, while improving balance of $U$ in both rough and smooth surfaces.  The method of \cite{Keele2015} for $\lambda = 0.382$ returned matches for which balance of the observed variables was better than for $\lambda = 0.05$, but $\lambda = 0.05$ returned matched data sets with better balance on $U$.

Lastly, the methods considered exhibited substantial variability in terms of the number of achieved matches. Optimal matching algorithms often return no matched pairs. Table \ref{table:dropped_units} shows the percentage of simulated data sets that each method failed to return any matches, and the average and IQR of number of treated units that were dropped when matching was achieved. GBM and matching in distance calipers had a high probability of failing to return matches, but when matching was achieved, they failed to match, on average, less than 1, or 0 treated units accordingly. DAPSm failed to return matches for 0.04\% of simulated data sets, but matched all treated units otherwise. On the other hand, \cite{Keele2015} returned matches with a significant amount ($\lambda = 0.05$) or a small number ($\lambda = 0.382$) of dropped treated units.  Differences in the number of obtained matches should be viewed in light of the fact that confining effect estimation to subsets of the available data can change the causal estimand of interest.

\begin{table}[t]
	\caption{Percentage of simulated data sets that each method returned no matches (\% fail), average number of treated units that were dropped when matches where returned (Dropped), the Interquartile range of number of dropped treated units (IQR), and average distance of matched pairs (Distance).}
	\label{table:dropped_units}
	\centering
	\footnotesize
	\begin{tabular}{lcccccccc}
		\hline
		& Gold PS & Na\"ive & N.Coords & GBM & DistCal 10\%  &  DAPSm  & Keele-0.05& Keele-0.382\\ 
		\hline
		\% fail & 1.5 & 0.5 & 0.96 & 27.67 & 33.29 & 0.04 & 0 & 0 \\
		Dropped & 0.06 & 0.06 & 0.06 & 0.23 & 0 & 0 & 55.98 & 2.11 \\ 
		IQR & (0,0) & (0,0) & (0,0) & (0,0) & (0,0) & (0,0) & (50,62) & (0,3) \\
		Distance ($\times 100$) & 37.4 & 40.5 & 36.4 & 36.1 & 8.4 & 2.6 & 1.9 & 3.7 \\
		\hline
	\end{tabular}
\end{table}


\section{Comparing the effectiveness of SCR/SNCR emission reduction technologies for reducing NO$_x$ emissions and ambient ozone}
\label{sec:data_app}

Regulatory strategies impacting U.S. power plants are predicated on the knowledge that reducing NO$_x$ emissions reduces ambient ozone, prompting many policies that incentivize the installation of emission control technologies at power plant smokestacks.   While many technologies are available, Selective Catalytic Reduction (SCR), and Selective Non-Catalytic Reduction (SNCR) technologies are believed to be among the most efficient for reducing NO$_x$.  However, no study has, to our knowledge, empirically compared the effectiveness of these strategies to evaluate whether the supposed efficiency gains of SCR/SNCR for reducing NO$_x$ emissions actually translate to greater reductions into ambient ozone concentrations.  


We compiled a national data source linking information on power plants, ambient pollution, population demographics, and weather.
The resulting data set consists of 473 power generating facilities powered by either coal or natural gas during June, July and August 2004, which represents the peak ozone season in a year following the institution of important NO$_x$ and ozone regulations. Covariate information ($\mathbf{X}$) on each facility includes power plant operating characteristics such as operating capacity and heat input, as well as area level characteristics such as temperature and population demographics.
As a measure of ozone in the area surrounding each power plant ($Y$), we use the fourth highest daily ozone concentration, averaged across all monitoring locations within a 100km radius.  This measure is chosen to mimic the National Ambient Air Quality Standard for ozone, which is based on the annual fourth-highest daily maximum 8-hour average ozone concentration.
Appendix \ref{sec:app_data_desc} has a detailed description on the exact construction of the data set used in the final analysis, including references to publicly-available raw data sets and R scripts used for data construction and linkage.

We consider as ``treated'' the power plants for which at least 50\% of the heat input is to facility units with at least one SCR or SNCR technology installed ($Z=1$, 152 facilities), with the remaining plants regarded as untreated ($Z=0$, 321 facilities). 67.7\% of facilities have either 0\% or 100\% of their heat input used by units with installed SCR or SNCR control technologies, suggesting robustness to the 50\% cutoff. Figure \ref{fig:data_map} shows maps of the power plants' treatment assignment and ozone measurements for the surrounding area, and Appendix \ref{sec:app_data_desc} discusses the emission control actions of the control group $Z = 0$.


To estimate the effect of SCR/SNCR technologies relative to alternatives, we implement the ``na\"ive'' approach, Matching within Distance Caliper (distance caliper was set to 354 miles, the $15^{th}$ percentile of all treated-control distances), GBM, the method of \cite{Keele2015}, and DAPSm (with standardized geo-distance). While the method of \cite{Keele2015} was implemented with the propensity score for the performance comparison in Section \ref{sec:sims}, here it is implemented in a manner more consistent with the intent of integer programming methods, matching directly on covariates.  The tolerance level for \cite{Keele2015} was set to 0.15 standard deviations for all continuous covariates.  Multiple values were tried for the tuning parameter $\lambda$, and results are presented with $\lambda =$ 800 ($41^{st}$ quantile of pairwise distances) such that the number of matched pairs will be similar to that of DAPSm. The caliper used for each method was decided such that methods would balance observed covariates, where ``balance'' is judged by an ASDM less than 0.15 (the same value used as the tolerance for the method of \cite{Keele2015}).

The variables that are included in the propensity score model are listed in Figure \ref{fig:DAPSw} and Appendix \ref{sec:app_variables}. Characteristics of the power plants (e.g. energy consumed, compliance scheme) are not expected to exhibit strong spatial patterns, but characteristics of the surrounding areas (e.g., temperature, population demographics) are.

\subsection{Covariate balance, number and distance of matched pairs}

\begin{figure}[!b]
	\centering
	\includegraphics[width = \textwidth]{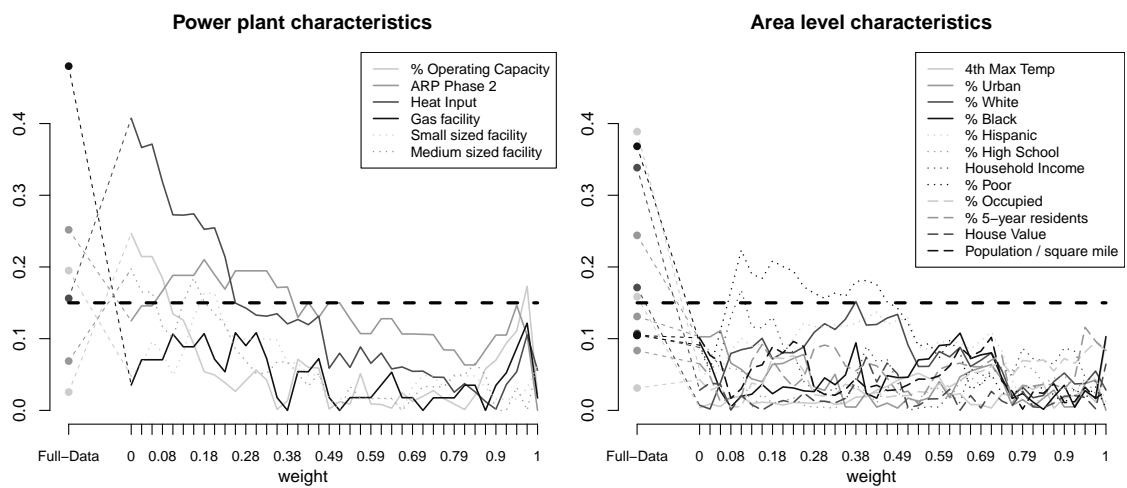} 
	\caption{Absolute standardized difference of means for covariates that are included in the propensity score model for the full data before any matching, and for various specification of the DAPSm weight. Balance of covariates on power plant characteristics is described on the left, and balance of area-level variables is shown on the right.}
	\label{fig:DAPSw}
\end{figure}

Covariate balance was assessed by comparing the covariate distribution of treated and control units.
Without adjustment, 10 out of 18 covariates were imbalanced between the treated and control facilities, as evidenced by the leftmost values of each panel in Figure \ref{fig:DAPSw}.  DAPSm was performed with values of $w$ ranging from 0 to 1, with covariate balance evaluated for each $w$, and depicted in the remaining portions of Figure \ref{fig:DAPSw}.  Note the change in covariate balance between the unadjusted setting and the setting with DAPSm($w=0$), which matches observations based solely on proximity.  Most area level characteristics achieve balance when matching only on proximity, but imbalance for power-plant level characteristics persists. Increasing values of $w$ place more emphasis on observed propensity score differences, and balance for covariates representing power-plant characteristics improves, without a strong sacrifice in balance for the area-level covariates. Using the procedure described in Section \ref{sec:choose_w}, $w\approx 0.513$ was chosen for the analysis.
Table \ref{table:ASDM} summarizes the covariate balance for all methods.  GBM failed to return balanced matched samples, and is excluded from the results.

\begin{figure}[p]
	\centering
	\includegraphics[width = 0.8\textwidth]{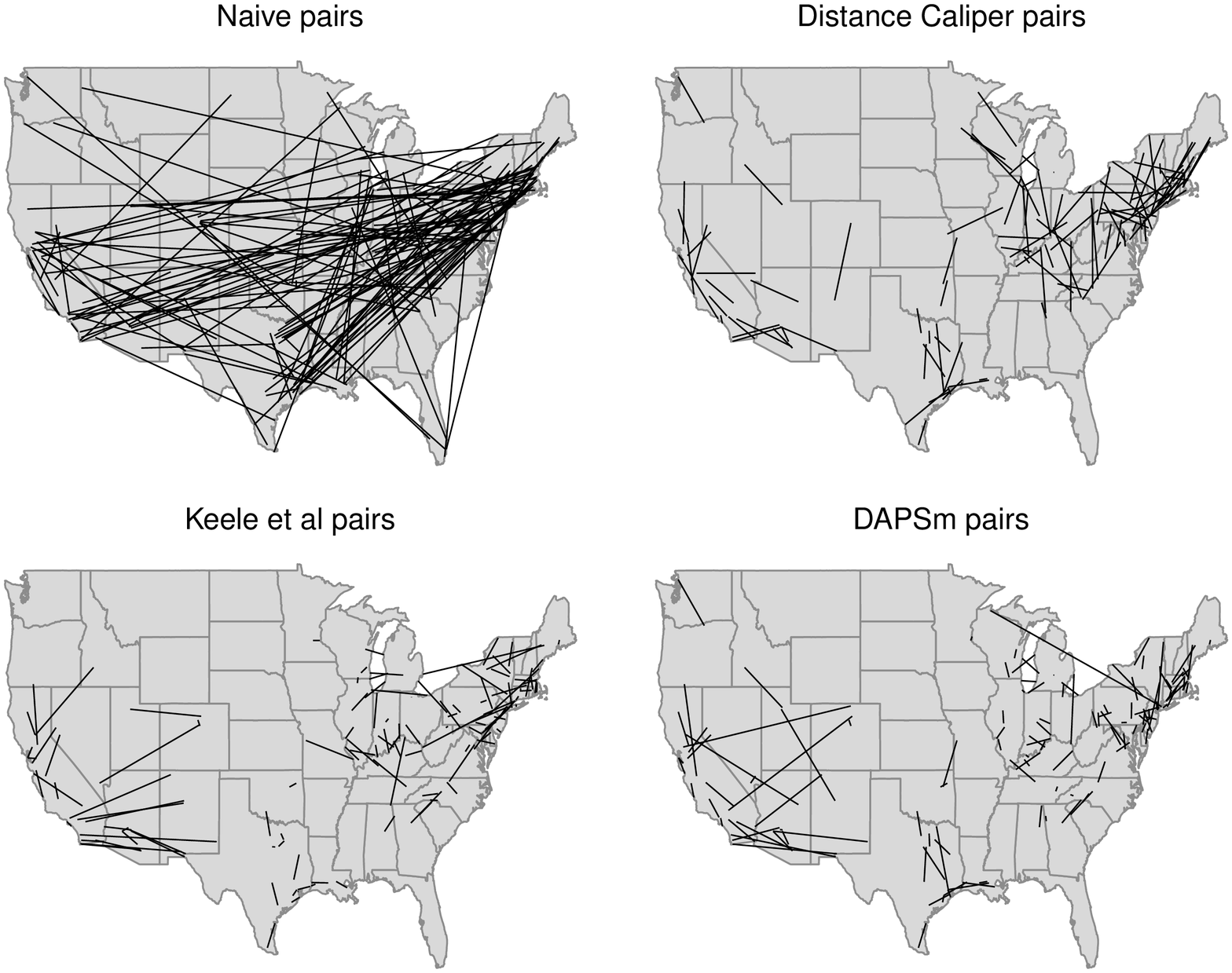}
	\caption{Maps of matched pairs for na\"ive, distance caliper, \cite{Keele2015}, and DAPSm approaches. Each line segment connects one treated power plant to its matched control.}
	\label{fig:matches}
	\vspace{20pt}
	
	\centering
	\includegraphics[width = \textwidth]{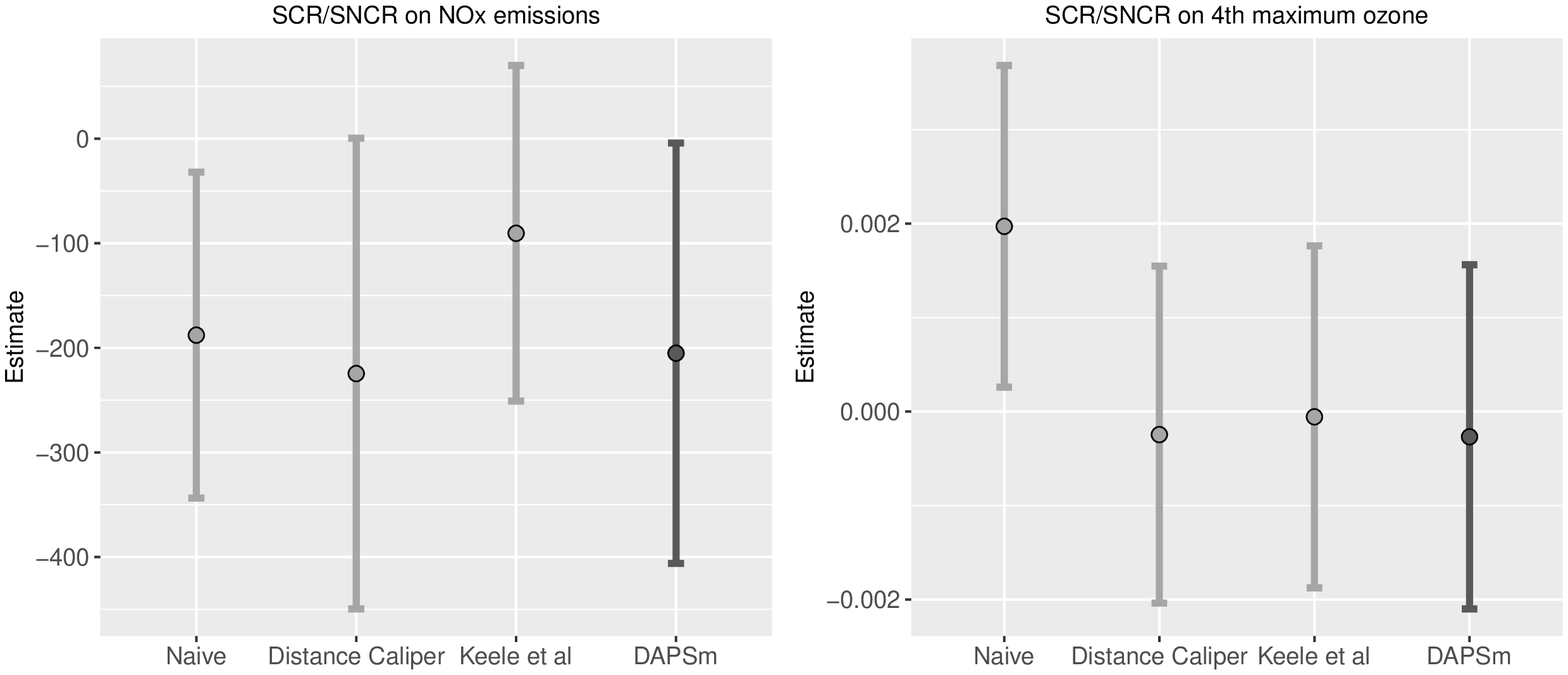}
	\caption{Effect estimates and 95\% confidence intervals for SCR/SNCR emission control technology installation on NO$_x$ emissions and 4$^{th}$ maximum ozone concentration during June-August 2004, using the na\"ive, Matching within Distance Caliper, \cite{Keele2015}, and DAPSm approaches.}
	\label{fig:app_results}
\end{figure}

Table \ref{table:ASDM} also presents the number and mean distance of matched pairs. Number of matches ranged between 116 and 137, indicating that not all of 152 treated units were matched.
Nonetheless, all methods should closely approximate the ATT in a manner that is comparable across methods, since most treated units are matched in all implementations. Characteristics of the matched population according to each method can be found in Appendix \ref{app_sec:matched_covs}. Dropped treated units were smaller, mostly gas-operating facilities in urban areas compared to the matched treated units.
Maps of the matched pairs are shown in Figure \ref{fig:matches}.

\begin{table}[t]
	\centering
	\caption{Balance of covariates assessed by the absolute standardized difference of means (ASDM)}
	\small
	\begin{tabular}{rrrrrr}
		\hline
		& Naive & Distance Caliper & Keele et al & DAPSm & Full-data \\ 
		\hline
		Number of imbalanced variables & 0 & 0 & 0 & 0 & 10 \\ 
		Mean ASDM & 0.067 & 0.052 & 0.065 & 0.045 & 0.189 \\ 
		Max ASDM & 0.148  & 0.134 & 0.145 & 0.150 & 0.480 \\ 
		Number of matches & 137 & 116 & 124 & 124 \\
		Mean distance (in miles) & 1066 & 198 & 146 & 141 \\
		\hline
	\end{tabular}
	\label{table:ASDM}
\end{table}

\subsection{Effect estimates for NO$_x$ and ozone}
We evaluate the effectiveness of SCR/SNCR technology for reducing NO$_x$ emissions and ambient ozone.  Since emissions are measured at the power plant, the analysis of NO$_x$ emissions is not expected to suffer from unmeasured spatial confounding. Since the formation of ambient ozone in the areas surrounding power plants is determined in part by atmospheric conditions, the analysis of ozone is expected to be susceptible to unmeasured spatial confounding. Confidence intervals are constructed conditional linear models fit to the matched data sets \citep{Ho2007}.  
Results from all methods are reported in Table \ref{table:app_results} and Figure \ref{fig:app_results}.

\subsubsection{Effects of SCR/SNCR on power plant NO$_x$ emissions}
Point estimates for the effect of SCR/SNCR on NO$_x$ emissions were below zero across all methods, with the na\"ive and DAPSm returning significant results at the 95\% confidence level. Power plants with installed SCR/SNCR emission control technologies emitted on average 205 tons of NO$_x$ less (95\% CI: 4 to 406 tons of NO$_x$ according to DAPSm) than what they would have had emitted had they adopted an alternative NO$_x$ control strategy.

\subsubsection{Effects of SCR/SNCR on ambient ozone}
In the analysis of ambient ozone concentrations, for which unmeasured spatial confounding is a concern, the na\"ive approach estimates a significant positive effect of SCR/SNCR installation on ambient ozone, which is inconsistent with the knowledge that SCR/SNCR reduces NO$_x$ emissions and the documented relationship between NO$_x$ and ozone.  This result corroborates suspicion of unmeasured confounding.
In contrast, estimates from all methods that incorporate spatial information provide estimates very close to zero (DAPSm: -0.27 parts per billion, 95\% CI: -2.1 - 1.56), indicating that SCR/SNCR does not reduce ambient ozone more than alternative strategies. For reference, these effect estimates can be compared against the national ozone air quality standard of 70 parts per billion.

\begin{table}[!b]
	\caption{Estimates and 95\% confidence intervals for the effect of SCR/SNCR on total NO$_x$ emissions (in tons) and $4^{th}$ maximum ozone measurement (in parts per billion).}
	\centering
	\begin{tabular}{rccccccc}
		\hline
		&    & NO$_x$ emissions & & & & Ozone & \\
		& LB & Estimate & UB & & LB & Estimate & UB\\ 
		\hline
		Naive 			 & -343.7 & -187.9 & -32.0  &&    0.26  & 1.97 & 3.68 \\ 
		Distance Caliper & -449.6 & -224.6 & 0.4    &&    -2.04 & -0.24 & 1.55 \\ 
		Keele et al 	 & -250.9 & -90.5  & 70 	&&    -1.88 & -0.06 & 1.76 \\ 
		\textbf{DAPSm} & \textbf{-406.1} & \textbf{-205.1} & \textbf{-4.1} &&      \textbf{-2.1} & \textbf{-0.27} & \textbf{1.56} \\ 
		\hline
	\end{tabular}
	\label{table:app_results}
\end{table}

\subsubsection{Comparison of effect estimates across methods}
As mentioned earlier, since NO$_x$ emissions are measured at the power plants' smokestacks, we do not expect unobserved spatial predictors of the outcome for this analysis. In fact, estimates across all  methods are similar.
    
However, in the analysis of ozone concentrations, we see that the spatial methods return results that are inconsistent with the na\"ive method. In Appendix \ref{sec:CE_w}, we provide additional evidence of the potential of unobserved spatial confounding in the analysis of ozone, by performing a sensitivity analysis of the DAPSm effect estimates as a function of $w$. The sensitivity analysis corroborates the existence of an unmeasured spatial confounder, with effect estimates that increase with $w$ (for $w > 0.513$) and approach the estimates from the na\"ive analysis when $w = 1$ and no adjustment for spatial proximity is made. In contrast, the sensitivity analysis of the effect of SCR/SNCR on NO$_x$ emissions indicates that spatial confounding is not an issue.

\section{Discussion}

Unobserved confounding is a ubiquitous issue in the analysis of observational studies. Settings with spatially-indexed data provide an opportunity to recover information on unobserved spatial confounding, but most methods have been confined to regression-based approaches. We propose a method that extends the benefits of propensity score matching procedures to settings with spatially-indexed data and provide a transparent and principled framework for assessing the relative trade offs of prioritizing observed confounding adjustment and spatial proximity adjustment.

The simulation study showed the potential for DAPSm to recover information on unobserved spatial confounding.  
When deployed to evaluate the effectiveness of emission control technologies, DAPSm balanced all observed covariates in the resulting matched data set while providing protection against the existence of unobserved spatial confounders.
The importance of incorporating spatial information was underscored by the ability of DAPSm (and other methods accounting for proximity) to return estimates that are more in line with subject-matter knowledge, in contrast to the na\"ive approach that ignores the possibility of  spatial confounding. Whereas the na\"ive approach indicated that clear reductions in NO$_x$ were accompanied by increases in ambient ozone, analysis with DAPSm (and other methods) provided the more credible result that SCR/SNCR do not decrease ambient ozone more than other strategies.

While we compare DAPSm against \cite{Keele2015} for illustration, it is improtant to remember the important fundamental distinction between these methods: DAPSm uses the propensity score while \cite{Keele2015} propose an integer programming method that matches on covariates directly. While a comparison between propensity score methods and integer programing methods is not the goal of this paper, it is worth noting that the most salient operational difference of the two methods relates to their respective tuning parameters that govern the amount of emphasis placed on matching observations that are geographically close.  DAPSm involves the tuning parameter ($w$) that offers a characterization of the price paid (in terms of observed covariate distance) by increasing emphasis on spatial proximity. This was evident in the ability to offer a practicable way to select a value of $w$ (as described in Section \ref{sec:choose_w}), and the transparent trade off between spatial proximity and observed covariate distance is an important feature of DAPSm that aligns with a scientific goal at the forefront of air pollution (and other) studies. On the other hand, the method of \cite{Keele2015} entails a tuning parameter ($\lambda$) that balances emphasis on spatial proximity against number of obtained matches for a fixed tolerance of covariate imbalance. \cite{Keele2015} provide an approach where, for a fixed tolerance, $\lambda$ could be chosen to obtain a target number of matches.  Further extensions growing from the mixed integer programing literature could give rise to alternative ways of prioritizing covariate balance, the number of matches, and the relative proximity of matches.

Furthermore, we evaluated the method of \cite{Keele2015} in simulations and in comparison with other reasonable approaches, in addition to DAPSm.  
These simulations showed that, across a variety of spatial confounding surfaces, DAPSm with an appropriately chosen $w$ performed comparably or better than the method of \cite{Keele2015} based on the propensity score, at least for some choices of $\lambda$.

While the comparison of different methods in the analysis of power plant emission controls highlights the potential for DAPSm to adjust for unmeasured spatial confounding, there are several important limitations to the analysis.  First, unmeasured (spatial or non-spatial) confounding may persist due to power plant or area level characteristics not contained in the data sources used. Second, we considered an ``active control'' group of power plants that did not install SCR/SNCR, but may have employed other strategies that could, in principle, be installed alongside SCR/SNCR (211 out of 311 control units employed a NO$_x$ control strategy other than SCR/SNCR). Finally, the analysis  relied on a simplification that linked each power plant to ambient ozone concentrations within a 100km radius. Importantly, this does not fully capture the phenomenon of long-range pollution transport whereby emissions from a particular source travel across large distances during conversion to ambient pollution.  Thus, installation of control technologies at a given power plant could affect ambient pollution concentrations around power plants located at distances greater than 100km, a phenomenon referred to as ``interference''.  While interference is not expected in the analysis  of NO$_x$ emissions, ignoring interference in the analysis of ozone concentrations has potential consequences.  The simplifications used here are expected to yield estimates that are closer to zero than any true effect of SCR/SNCR on ambient ozone, as installation of these technologies is likely to reduce ambient ozone even around power plants that were considered in the ``control group'' for this analysis.  Methods for causal inference with interference have been recently considered with spatially-indexed data \citep{Verbitsky-savitz2012, Zigler2012}, including our own current work on methods advances to address interference in this specific setting \citep{Papadogeorgou2017}.  Furthermore, the analysis relies on some extent on correct specification of the propensity score model, and \cite{King2016} argue against the use of propensity score for matching altogether. For that reason, checking covariate balance in the design phase \citep{Rubin2008} without evaluating outcomes, is an important component of propensity score matching.

\section*{Acknowledgments}

Funding for this work was provided by National Institutes of Health R01ES026217, P01CA134294, R01GM111339. USEPA 83587201-0, and Health Effects Institute 4953-RFA14-3/16-4. The contents of this work are solely the responsibility of the grantee and do not necessarily represent the official views of the USEPA. Further, USEPA does not endorse the purchase of any commercial products or services mentioned in the publication.

\bibliographystyle{apalike}
\bibliography{Spatial_PS}

\appendix


\setcounter{figure}{0}    
\numberwithin{figure}{section}
\numberwithin{equation}{section}
\numberwithin{table}{section}

\section{Alternative definition of standardized distance}
\label{app_sec:sd2}
Another specification of the distance measure in DAPS could be the empirical CDF of all treated-control pairwise distances. Using this definition, treated unit $i$ and control unit $j$ have distance defined as:
\begin{equation}
Dist_{ij} = \frac{\sum_{k \in S_t, l \in S_c }\mathbb{I}\{d_{kl} \leq d_{ij} \}}{N_t \times N_c}.
\tag{A.1}
\label{eq:SD2}
\end{equation}
With both definitions of the distance measure, $Dist_{ij} \in [0, 1]$ for every $i$ treated, and $j$ control.

In the simulations, the performance of DAPSm was similar with respect to MSE, absolute bias, number of matches, and balance of observed and unobserved covariates for the two specifications of standardized distance. In general, specifying distance as in (\ref{eq:SD2}) led to a smaller $w$ chosen than the specification of (\ref{eq:SD1}). However, this is expected when examining the relationship between the two distance measures.

\section{Greedy DAPSm algorithm}
\label{app_sec:greedy}
Consider the DAPS table of all pairs defined as:
\begin{table}[H]
	\centering
	\footnotesize
	\caption{The matrix of calculated DAPS for treated-control pairs.}
	\begin{tabular}{|cc|cccc|}
		\hline
		& & & Controls & & \\
		& & 1 & 2 & \dots & N$_c$ \\[5pt]
		\hline
		& 1 & $DAPS_{11}$ & $DAPS_{12}$ & \dots & $DAPS_{1N_c}$ \\[5pt]
		Treated & 2 & $DAPS_{21}$ & $DAPS_{22}$ & \dots & $DAPS_{2N_c}$ \\[5pt]
		& \vdots &&&\vdots& \\[5pt]
		& N$_t$ & $DAPS_{N_t1}$ & $DAPS_{N_t2}$ & \dots & $DAPS_{N_tN_c}$\\
		\hline
	\end{tabular}
\end{table}
\normalsize
Entries are set to infinity if the DAPS value of the pair is larger than the caliper.

The minimum element of each row is identified. These minimum values correspond to the minimum distance to a control for every treated unit, and rows with minimum value equal to infinity are dropped (treated units without any control within the caliper). The matrix is reordered in increasing order of these minimum values, and the controls that acheive them are identified for every treated unit.

If there is no overlap in the control units, all treated units are matched to the controls with minimum distance. Otherwise, treated units are matched up to the first control that is repeated. In that case, the new minimum distance values are calculated over the rows of the new matrix (with matched rows and columns dropped), and the procedure is repeated.


For data sets large enough to preclude the practicality of employing DAPSm with many values of $w$, an algorithm which is based explicitly on assuming a non-increasing trend of ASDM with $w$ and uses fewer fits of DAPSm can be employed.
The procedure is initiated at $w = 0.5$ (step $k=1$). If balance is achieved at step $k - 1$, $w$ is decreased by $1/2^{k + 1}$ and balance is re-accessed. If balance is not achieved at step $k - 1$, $w$ is increased by $1/2^{k + 1}$. The procedure is iterated and it ends at the optimal $w$ when the step size is smaller than a pre-specified tolerance level.

\section{Data generating mechanism for simulation study}
\label{app:sims_generate}

For every pair of $\nu, r$ of the spatial variable, we simulate 100 data sets. For each data set, each of 800 fixed locations are simulated to have:
\begin{enumerate}
	\item One unmeasured confounder, $U$, generated as a Gaussian Process with a Mat\'ern($\nu, r$) correlation function, normalized to have mean 0 and variance 1.
	\item Four observed confounders, $X_i$, simulated as independent normal variables with mean 0, variance 1, and $Cor(U, X_i) = 0$, $i = 1, 2, 3, 4$.
	\item
	Treatment, $Z$, generated as a binary variable with
	$$
	\mathrm{logit} P(Z = 1) = - 0.85  + 0.1\ X_1 + 0.2\ X_2 -0.1\ X_3 -0.1\ X_4 + 0.3\ U
	$$
	This generative model for the treatment gives rise to data sets with approximately 30\% of observations treated.
	
	\item Outcome, $Y$, generated as:
	$$
	Y =  \ Z + 0.55\ X_1 + 0.21\ X_2 + 1.17\ X_3 -0.11\ X_4 + 3\ U + \epsilon, \ \epsilon \sim  N(0, 1)
	$$
\end{enumerate}

\newpage
\section{Surfaces of Mat\'ern spatial variable}
\label{fig:Usurf}
$ \ $
\vspace{-30pt}
\begin{figure}[H]
	\centering
	\includegraphics[width=0.8\textwidth]{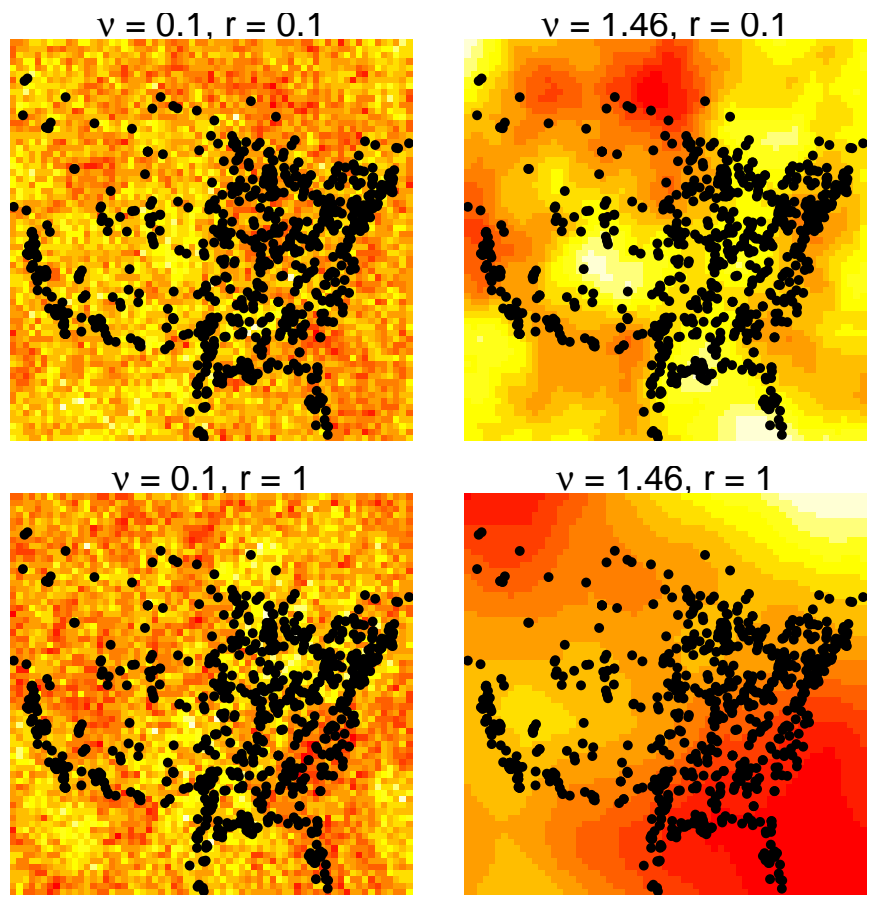}
	\caption{Surfaces of spatial variables with Mat\'ern correlation function with smoothness $\nu$ and range $r$. Points represent 800 power plant locations.}
\end{figure}

\section{Simulation results for the method of \cite{Keele2015} matching directly on covariates}
\label{app_sec:sims_Keele}

In section \ref{sec:sims} of the main text, we implement a short simulation study to examine the performance of various methods to incorporate spatial information in a matching procedure.
Most methods considered use the propensity score to adjust for observed confounders. In the simulation study of the main text, the method of \cite{Keele2015} was implemented by matching exactly on 5 categories of the propensity score for comparison.

However, \cite{Keele2015} propose their method within an integer-programming context and argue for matching directly on covariates using, for example, moment matching on continuous covariates and exact matching on discrete variables. Here, we present the relative MSE of \cite{Keele2015} for matching on covariates $X_1, X_2, X_3, X_4$ using moment matching with tolerance equal to $0.1$ standard deviations, and a few specifications of the parameter $\lambda$.

\begin{figure}[H]
	\includegraphics[width = 0.9\textwidth]{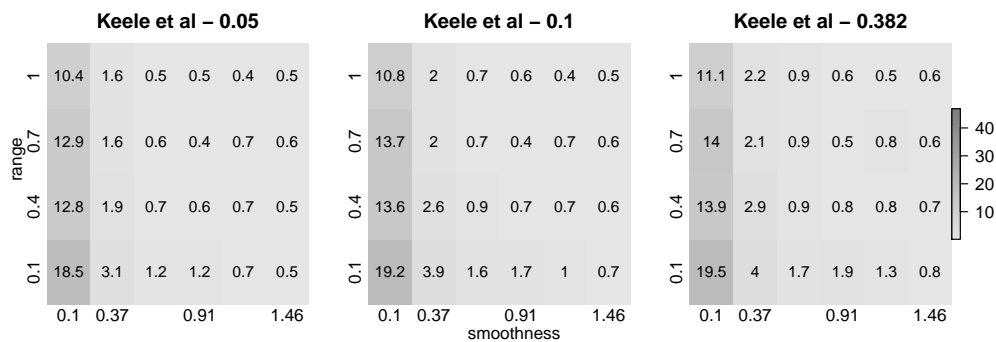}
	\caption{MSE of the method of \cite{Keele2015} with respect to the gold standard propensity score matching approach. Moment matching was performed for the continuous observed variables with tolerance set to 0.1 standard deviation.}
\end{figure}

The mean (IQR) number of dropped treated units for values of $\lambda \in \{0.05, 0.1, 0.382\}$ was 11.7 (8,15), 2.2 (0,3), and 0.02 (0,0) accordingly.

\section{Constructing the analysis data set}
\label{sec:app_data_desc}

All tools and data sets required to construct the analysis data set are publicly available and easily accessible. They include:
\begin{enumerate}
	\item Raw data files (AMPD-EIA on power plants, AQS temperature data, AQS ozone data, Census 2000 data) available at \url{https://dataverse.harvard.edu/dataverse/dapsm}.
	\item The DAPSm R Package (available at \url{https://github.com/gpapadog/DAPSm})
	\item The AREPA R Package (available at \url{https://github.com/czigler/arepa})
	\item R scripts that perform the data manipulation and linking of the raw data sets (available at \url{https://github.com/gpapadog/DAPSm-Analysis}).
\end{enumerate}

A power plant can consist of more than one energy generating unit (EGU). We restrict our analysis to power generating units that are using coal or natural gas as one of their primary fuels. Power plant covariate information is monthly and measured for each of its EGUs. EGUs that were retired, not operating, not yet operating during June, July and August of 2004, or did not have any data before or started having data after this time period were dropped.

A key covariate measured at the EGU level is heat input describing the energy used by the power plant unit for operation, and is therefore a good predictor of its size. Over the period of three months (June, July, August 2004), 14\% of all entries lack heat input information. 82\% of the missing data was imputed using heat input information from other years (2003, 2002, 2005, 2006) (R squared of the models used ranges between 89.8-94.7\%). 25\% of predicted heat input observations were estimated to close to zero but negative.  These values were set to 0. Monthly unit level data were afterwards aggregated to the facility level and over the June-August 2004 period. 68 (5\%) facilities were dropped because of missing heat input. 

For each ozone monitoring site, temperature information was assigned as the average temperature over all temperature monitoring sites within 150 kilometers. Population demographics for the year 2000 for the surrounding area were assigned to each ozone monitor as aggregated zip code-level Census variables with centroids located within 6 miles of the monitor. The resulting data set includes ozone, temperature and demographics measured at or around ozone monitoring sites.

Finally, ozone, temperature and population demographics were assigned to power plants as the average over ozone monitors within 100 kilometers. Each monitor was only allowed to contribute to the closest power plant. A monitor site is not linked to any power plant if there is no power plant within 100 kilometers. A power plant is not linked to any ozone monitor, if there is no ozone monitor within 100 kilometers without another power plant  closer.

The resulting data set consists of 483 power plants linked to a total of 937 ozone monitors. 10 additional facilities are dropped due to missing Census information, or missing percent capacity, resulting to 473 facilities in our final data set linked to a total of 921 ozone monitoring sites.

A facility is considered treated if at least 50\% of its heat input is used by EGUs with at least one SCR/SNCR installed. 1,230 out of 2,964 EGUs in control facilities have no NO$_x$ emission control technologies installed, while 33 have one of SCR/SNCR installed (amounting to less than 50\% of the facility's heat input). The remaining units that constitute the facilities in the control group have some other type of NO$_x$ control installed such as a low NO$_x$ burner, an overfire reduction system, an ammonia injection system, a modified combustion method, water injection system, or some other non-specified control.  All of these alternative strategies in the control group are designed to reduce NO$_x$, but are widely regarded as less efficient than SCR/SNCR.
Note that it is very common for EGUs (treated or control) to follow more than one NO$_x$ emission control strategies, implying that even EGUs in the control group having other NO$_x$ control strategies might still be candidates for additional installation of SCR/SNCR. 

\section{Data application covariate description}
\label{sec:app_variables}
$\ $
{
	\footnotesize
	\begin{longtable}{L{1.7cm}  L{3.7cm}  c    ccc}
		\caption{Power plant and area level characteristics before matching}
		\label{table:app_variables}\\
		Name & Description & Units & Treated & Control & ASDM  \\
		\hline
		\% Capacity & Percentage of operating capacity & - & 0.42 (0.28) & 0.42 (0.3) & 0.026 \\
		Heat Input & Amount of fuel energy burned for power generation (logarithm) & log MMBtu & 14.3 (1.97) & 14 (1.96) & 0.156 \\
		Phase 2 & ARP Phase 2 indicator & - & 0.86 (0.35) &  0.77 (0.42) & 0.252 \\
		Gas & Mostly gas burning power plant & - &  0.77 (0.42) & 0.57 (0.5) & 0.48 \\
		Small sized & Power plants consisted by 1 or 2 EGUs & - & 0.62 (0.49) & 0.52 (0.5) & 0.195 \\
		Medium sized & Power plants consisted by 3, 4, or 5 EGUs & - & 0.33 (0.47) & 0.36 (0.48) & -0.069 \\
		\hline
		Temperature & 4$^{th}$ maximum temperature over study period & Fahrenheit & 70.7 (8.1) & 69.5 (7.7) &  0.158 \\
		\% Urban & Percentage of population in urban areas & - & 0.76 (0.32) & 0.72 (0.34) & 0.131 \\
		\% White & Percentage of white population & - & 0.77 (0.17) & 0.80 (0.17) & -0.171 \\
		\% Black & Percentage of black population & - & 0.08 (0.09) & 0.11 (0.14) & -0.368 \\
		\% Hispanic & Percentage of hispanic population & - & 0.16 (0.18) & 0.09 (0.14) & 0.389 \\
		\% High School & Percentage of population that attended high school & - & 0.29 (0.08) & 0.31 (0.08) & -0.244 \\
		Household Income & Median household income & USD & 44,721 (12,456) & 43,386 (12,223) & 0.107 \\
		\% Poor & Percentage of impoverished population & - & 0.13 (0.06) & 0.12 (0.06) & 0.105 \\
		\% Occupied & Percentage of occupied population  & - & 0.91 (0.11) & 0.91 (0.1) & 0.031 \\
		\% MovedIn5 & Percentage of population that has lived in the area for less than 5 years & - & 0.48 (0.08) & 0.47 (0.09) & 0.083 \\
		House value & Median house value & USD & 148,394 (97,144) & 115,510 (57,472) & 0.339 \\
		Population density & Population per square mile (logarithm) & log \# / mile$^2$ & 6.19 (1.68) & 6.02 (1.71) & 0.105 \\
		\hline
	\end{longtable}
}
\section{Description of matched and dropped treated units}
\label{app_sec:matched_covs}
In Table \ref{app_tab:matched_covs} we show the mean and standard deviation of the matched and dropped treated units for each method. The mean and standard deviation of the treated units in the full data can be found in Table \ref{table:app_variables}.

\footnotesize
\begin{longtable}{L{3cm}  cccc}
	\caption{Covariate mean and standard deviation for matched and dropped treated units for the Na\"ive, Matching in Distance Caliper, \cite{Keele2015}, and DAPSm.}
	\label{app_tab:matched_covs}\\
	& Na\"ive & Distance Caliper & \cite{Keele2015} & DAPSm \\
	\hline
	\textbf{Matched units} \\
	\% Capacity 	& 0.43 (0.29)   & 0.4  (0.27)   & 0.36 (0.27)   & 0.42 (0.28) \\
	Heat Input 		& 14.4 (1.96)   & 14.1 (1.98)   & 13.8 (2.07)   & 14.3 (2.05) \\
	Phase 2 		& 0.85 (0.36)   & 0.86 (0.35)   & 0.91 (0.29)   & 0.82 (0.38) \\
	Gas 			& 0.69 (0.46)   & 0.81 (0.39)   & 0.85 (0.36)   & 0.73 (0.45) \\
	Small sized 	& 0.58 (0.5)    & 0.72 (0.45)   & 0.70 (0.46)   & 0.59 (0.49) \\
	Medium sized 	& 0.36 (0.48)   & 0.25 (0.43)   & 0.26 (0.44)   & 0.36 (0.48)  \\
	Temperature 	& 70.1 (7.59)   & 70.8 (7.79)   & 71.5 (7.72)   & 70.6 (8.59) \\
	\% Urban 		& 0.79 (0.29)   & 0.78 (0.31)   & 0.79 (0.3)    & 0.72 (0.34) \\
	\% White 		& 0.77 (0.17)   & 0.76 (0.17)   & 0.75 (0.18)   & 0.79 (0.16)\\
	\% Black 		& 0.08 (0.1)    & 0.06 (0.07)   & 0.06  (0.06)  & 0.08 (0.1) \\
	\% Hispanic  	& 0.15 (0.19)   & 0.18 (0.19)   & 0.20 (0.21)   & 0.13 (0.17) \\
	\% High School 	& 0.30 (0.08)   & 0.28 (0.08)   & 0.28 (0.08)   & 0.31 (0.08) \\
	Household Income& 44166 (10964) & 45430 (11994) & 45846 (11748) & 44496 (12654)\\
	\% Poor 		& 0.12 (0.06)   & 0.13  (0.06)  & 0.13 (0.06)   & 0.12 (0.06)\\
	\% Occupied 	& 0.92 (0.05)   & 0.93  (0.05)  & 0.92 (0.06)   & 0.90 (0.12)\\
	\% MovedIn5 	& 0.48 (0.07)   & 0.49  (0.07)  & 0.49 (0.07)   & 0.47 (0.08)\\
	House value 	& 142750 (79501)& 155577 (83956)& 161888 (86553) & 137894 (98511)\\
	\hline
	\textbf{Dropped units} \\
	\% Capacity 	& 0.42 (0.27) & 0.44 (0.28) & 0.47 (0.28) & 0.44 (0.25) \\ 
	Heat Input 		& 14.25 (2) & 14.51 (1.96) & 14.69 (1.83) & 14.25 (1.63) \\ 
	Phase 2 		& 0.87 (0.34) & 0.86 (0.35) & 0.81 (0.39) & 1 (0) \\ 
	Gas 			& 0.87 (0.34) & 0.73 (0.44) & 0.71 (0.46) & 0.96 (0.19) \\ 
	Small sized 	& 0.67 (0.47) & 0.53 (0.5) & 0.56 (0.5) & 0.75 (0.44) \\ 
	Medium sized 	& 0.28 (0.45) & 0.4 (0.49) & 0.38 (0.49) & 0.18 (0.39) \\ 
	Temperature 	& 71.5 (8.7) & 70.7 (8.39) & 70.19 (8.38) & 71.52 (5.46) \\ 
	\% Urban 		& 0.72 (0.35) & 0.75 (0.32) & 0.74 (0.33) & 0.92 (0.11) \\ 
	\% White 		& 0.77 (0.17) & 0.78 (0.17) & 0.78 (0.16) & 0.65 (0.16) \\ 
	\% Black 		& 0.07 (0.09) & 0.09 (0.11) & 0.1 (0.11) & 0.06 (0.07) \\ 
	\% Hispanic  	& 0.17 (0.17) & 0.14 (0.18) & 0.13 (0.16) & 0.29 (0.17) \\ 
	\% High School 	& 0.28 (0.08) & 0.3 (0.08) & 0.31 (0.08) & 0.23 (0.04) \\ 
	Household Income& 45425 (14180) & 44131 (12871) & 43858 (12976) & 45719 (11702) \\ 
	\% Poor 		& 0.13 (0.06) & 0.13 (0.06) & 0.13 (0.06) & 0.14 (0.05) \\ 
	\% Occupied 	& 0.89 (0.15) & 0.9 (0.14) & 0.9 (0.13) & 0.95 (0.02) \\ 
	\% MovedIn5 	& 0.48 (0.08) & 0.48 (0.08) & 0.47 (0.08) & 0.53 (0.04) \\ 
	House value 	& 155557 (115989) & 142424 (107020) & 138040 (103856) & 194898 (76283) \\ 
	\hline
\end{longtable}
%
%
%

\section{DAPSm effect estimates as a function of the tuning parameter}
\label{sec:CE_w}

\begin{figure}[!b]
	\centering
	\includegraphics[width = \textwidth]{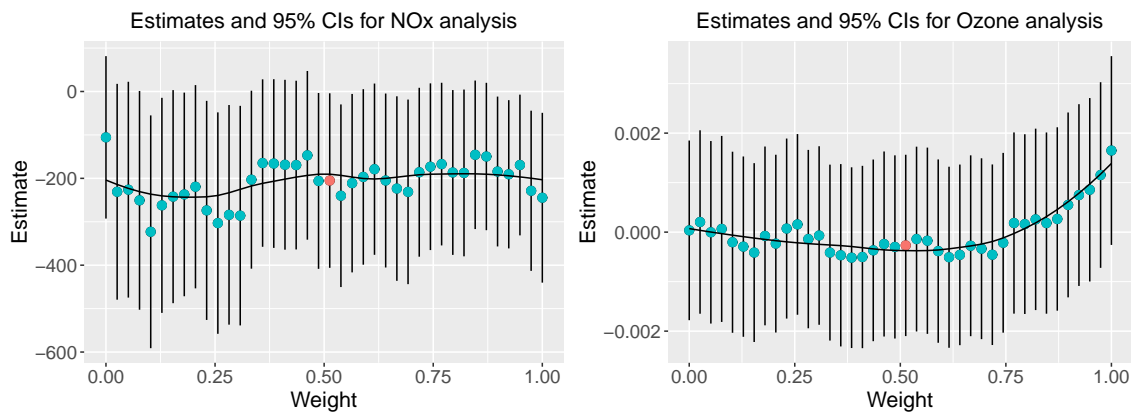}
	\caption{Estimates and 95\% confidence intervals using DAPSm as a function of $w$ for the NO$_x$ emissions (left) and O$_3$ (right)  analyses. The red point corresponds to the effect estimate based on $w = 0.513$.}
	\label{fig:CE_w}
\end{figure}

\begin{figure}[t]
	\centering
	\includegraphics[width = 0.55\textwidth]{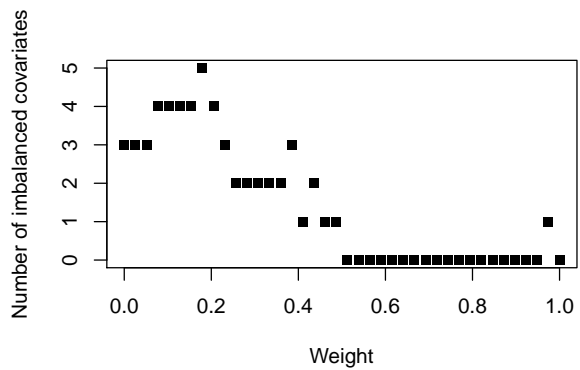}
	\caption{Number of imbalanced variables as a function of $w$ in DAPSm.}
	\label{fig:imbalance}
\end{figure}

We investigate the sensitivity of the DAPSm results to the specification of $w$, by plotting the estimates and 95\% confidence intervals of the ``effect'' estimates as a function of $w$. First, note the number of imbalanced covariates is generally decreasing as a function of $w$ (Figure \ref{fig:imbalance}), indicating that a weight over the chosen 0.513 is necessary to balance all observed covariates. Therefore, the estimates for small values of $w$ are not necessarily interpretable, since residual confounding might still be present.     
As the value of $w$ increase above the chosen value, balance of observed covariates is almost always maintained, while less weight is given to achieving matches at close proximity. As $w$ tends to 1, DAPSm resembles simple propensity score matching and incorporates a decreased amount of spatial information.

As mentioned on the main text, unobserved spatial confounding is unlikely to be present in the analysis of NO$_x$ emissions, since NO$_x$ emissions are measured directly at the plant's smokestacks, and area-level characteristics are unlikely to be predictors. In the left panel of Figure \ref{fig:CE_w}, we see that the effect estimates for the NO$_x$ analysis remain more or less constant for values of $w \geq 0.513$, indicating that as long as observed covariates are balanced, the effect estimates will be similar and independent of the chosen value of $w$.

However, in the right panel of Figure \ref{fig:CE_w}, we see a very different result. Specifically, the effect estimates are increasing for values of $w$ greater than 0.513. This might imply that, when observed covariates are balanced, matching on spatial proximity is an important component of acquiring unbiased effect estimates. Specifically, it indicates that there might be an unobserved spatial confounder, which leads to positive bias when it is not adjusted for.

\end{document}